\input{aipcheck}

\documentclass[final,numberedheadings,letterpaper]{aipproc}
\usepackage{color}
\usepackage{graphicx}
\usepackage{todonotes}

\layoutstyle{6x9}

\newcommand{\Eref}[1]{{Eq.~(\ref{#1})}}
\newcommand{\Fref}[1]{{Fig.~\ref{#1}}}

\begin{document}
\title{Fidelity decay of the two-level bosonic embedded ensembles of Random Matrices}



\author{Luis Benet}{
  address={Instituto de Ciencias F\'isicas, Universidad Nacional Aut\'onoma de M\'exico,
  Cuernavaca, M\'exico}
}

\author{Sa\'ul Hern\'andez-Quiroz}{
  address={Instituto de Ciencias F\'isicas, Universidad Nacional Aut\'onoma de M\'exico,
  Cuernavaca, M\'exico},
  altaddress={Facultad de Ciencias, Universidad Aut\'onoma del Estado de Morelos, 
  Cuernavaca, M\'exico}
}

\author{Thomas H. Seligman}{
  address={Instituto de Ciencias F\'isicas, Universidad Nacional Aut\'onoma de M\'exico,
  Cuernavaca, M\'exico},
  altaddress={Centro Internacional de Ciencias, Cuernavaca, M\'exico} 
}

\begin{abstract}
We study the fidelity decay of the $k$-body embedded ensembles of random 
matrices for bosons distributed over two single-particle states. Fidelity is defined in
terms of a reference Hamiltonian, which is a purely diagonal matrix consisting of 
a fixed one-body term and includes the diagonal of the perturbing $k$-body embedded 
ensemble matrix, and the perturbed Hamiltonian which includes the residual 
off-diagonal elements of the $k$-body interaction. This choice mimics the typical 
mean-field basis used in many calculations. We study separately the cases 
$k=2$ and $3$. We compute the ensemble-averaged fidelity decay as well as the fidelity 
of typical members with respect to an initial random state. Average fidelity displays 
a revival at the Heisenberg time, $t=t_H=1$, and a freeze in the fidelity decay, during 
which periodic revivals of period $t_H$ are observed. We obtain the relevant scaling 
properties with respect to the number of bosons and the strength of the perturbation. For 
certain members of the ensemble, we find that the period of the revivals during the 
freeze of fidelity occurs at fractional times of $t_H$. These fractional periodic revivals 
are related to the dominance of specific $k$-body terms in the perturbation. 
\end{abstract}

\maketitle

\section{Introduction}
Fidelity, also named quantum Loschmidt echo, is a measure of the sensitivity 
of the dynamics of quantum systems to perturbations and has attracted 
a lot of attention in the last years in connection with quantum information processes
and quantum chaology; for a recent review see~\cite{Gorin2006}. 

Fidelity compares the time evolution of a given initial state under a reference 
Hamiltonian with the time evolution of the same state under a slightly different one. 
We denote as ${\hat H}_0$ the reference or unperturbed Hamiltonian, 
and as ${\hat H}_\epsilon = {\hat H}_0+\epsilon {\hat V}$ the perturbed one which 
includes the residual interaction. The former is associated with the unitary
time-evolution ${\cal U}_0(t)=\hat{T}\exp[-\frac{i}{\hbar}\int_0^t d\tau\, {\hat H}_0(\tau)]$, 
where ${\hat T}$ is the time-ordering operator; likewise, ${\cal U}_\epsilon(t)$ is the 
propagator associated with the perturbed Hamiltonian. In the definitions of the
propagators we have considered the most general case in which the Hamiltonians may 
depend on $t$ explicitly. With these definitions, and taking an arbitrary initial state 
$|\Psi_0\rangle$, the fidelity amplitude is defined as
\begin{equation}
\label{eq1}
f_\epsilon(t) = \langle \Psi_0| {\cal U}_0(-t) {\cal U}_\epsilon(t) |\Psi_0\rangle,
\end{equation}
whose square modulus is known as the {\it fidelity}
\begin{equation}
\label{eq2}
F_\epsilon(t) = |f_\epsilon(t)|^2.
\end{equation}

Fidelity decay can occur in many different ways and in successive combinations of 
these. For an extremely short time, the so-called Zeno time, all systems will decay 
quadratically in time, both for chaotic and integrable systems. After the Zeno time 
systems specific correlations often dominate the decay.

For regular systems the decay typically changes into a quadratic and later Gaussian 
decay dominated by the dimension of the effective Hilbert space that is accessible to 
the wave function. Furthermore revivals are very common, particularly if the system 
has few degrees of freedom.

For chaotic systems on the other hand we expect, after the system specific behavior, 
linear or exponential decay of the fidelity which is determined by the 
coupling strength. This phase is often known as Fermi Golden Rule decay.
Around the Heisenberg time the exponential decay changes over into a quadratic 
function or a Gaussian. An alternative behavior, known as Lyapunov decay, can occur 
with rather strong perturbations. For times of the order of the Ehrenfest time, exponential 
decay can be observed. The Lyapunov exponent, and not the perturbation strength, define 
the decay rate.

An alternative and interesting behavior is found if the diagonal part of the perturbation 
(or at least its time-average) vanish in the basis in which the unperturbed Hamiltonian
is diagonal. Then we get a fidelity freeze, which has been shown for classically 
integrable~\cite{ProsenZnid2003} and chaotic~\cite{ProsenZnid2005} systems, 
as well as for random matrix models~\cite{GKPSSZ2006}. Both, the value at which the 
freeze occurs and the length of its duration, are functions of perturbation strength. For 
fermions, it was proposed that a mean field Hamiltonian as the reference system and 
the residual interaction could well lead to a fidelity freeze~\cite{Iztok2007}. Specifically, 
this problem was treated in the framework of an embedded two-body random ensemble
for fermions~\cite{BW2003}. While there is no freeze on average, because of long tails 
in the distribution, typical samples from the ensemble display the freeze and indeed 
the median does~\cite{Iztok2007}.

In this paper we study the fidelity decay for the $k$-body embedded ensembles
of random matrices for bosons, which are distributed over two single-particle 
states. We shall consider a reference Hamiltonian which consists 
purely of a diagonal $1$-body terms as well as the diagonal $k$-body matrix elements,
with $k=2$ or $k=3$, which are the physical important cases. The perturbed Hamiltonian 
will include a traceless residual interaction, consisting of the off-diagonal matrix elements 
of the same $k$-body interaction. Our results indicate that ensemble-averaged 
fidelity decays quadratically on time for short times (before the Heisenberg time $t_H$) 
as well as for very long times, precisely after the freeze of fidelity ends. Furthermore, 
freeze of fidelity is observed for the ensemble--averaged fidelity, and 
during the freeze, it displays periodic revivals with the periodicity given by 
the Heisenberg time. These results occur {\it often but not always,} when individual 
members of the ensemble are considered. We present numerical results 
showing that the periodicity of the revivals is an integer fraction of $t_H$, and 
demonstrate that the specific value of this period is related with specific off-diagonal
terms of the $k$-body interaction considered in the perturbation.

\section{The $k$-body two-level bosonic ensemble of random matrices}

The $k$-body Embedded Ensemble of Random Matrices for bosons considers
all possible $k$-body interactions among spin-less $n$-boson states, where the bosons 
are distributed over $l$ single-particle states~\cite{Asaga2001}. Below 
we discuss the simplest case $l=2$. To define this ensemble, we first introduce the 
single-particle states associated with the operators $\hat a_j^\dagger$ and 
$\hat a_j$, with $j=1,2$ ($l=2$), which, respectively, create or annihilate one boson 
on the single-particle level $j$. These operators satisfy the usual commutation 
relations for bosons, i.e., $[\hat{a}_i,\hat{a}_j^\dagger]=\delta_{ij}$. The 
normalized $n$-boson states are denoted as 
$|\mu_r^{(n)}\rangle = ({\cal N}_r^{(n)})^{-1} (\hat a_1^\dagger)^r (\hat
a_2^\dagger)^{n-r}|0\rangle$, where ${\cal N}_r^{(n)}=[r!
(n-r)!]^{1/2}$ is a normalization constant and $|0\rangle$ is the
vacuum state. The Hilbert--space dimension is thus $N=n+1$.
Then, in second-quantized form, the most general $k$-body interaction of 
$n$ bosons distributed over two single-particle levels, $\hat H_k^{(\beta)}$, 
can be written as~\cite{BLS2003}
\begin{equation}
  \label{eq3}
  {\hat H_k^{(\beta)}} = \sum_{r,s=0}^k \, v_{r,s}^{(\beta)} \,
  \frac{ (\hat a_1^\dagger )^{r} ( \hat a_2^\dagger )^{k-r} 
    (\hat a_1 )^{s} ( \hat a_2 )^{k-s} }{{\cal N}_r^{(k)} {\cal N}_s^{(k)}}  \ .
\end{equation}
Here, $k$ denotes the rank of the interaction, $1\le k \le n$, and $v_{r,s}^{(\beta)}$
are the $k$-body matrix elements, which are independent Gaussian-distributed 
random numbers with zero mean and constant (fixed) variance $v_0^2=1$. As in the case 
of the canonical random matrix ensembles (see~\cite{GMGW1998}), Dyson's parameter 
$\beta$ distinguishes the cases according to time-reversal invariance: 
$\beta=1$ is the case where time-reversal invariance holds, and 
$\beta=2$ where this symmetry is broken. Hence, the $k$-body interaction matrix 
$v^{(\beta)}$ is a member of the Gaussian orthogonal ensemble (GOE) for
$\beta=1$, or Gaussian unitary ensemble (GUE) for $\beta=2$. 
Note that $\hat{H}_k^{(\beta)}$ commutes with the number operator 
${\hat n}=a_1^\dagger a_1+a_2^\dagger a_2$, i.e., the interaction preserves the 
total number of bosons $n$. 

This ensemble presents some noteworthy properties: It exhibits non-ergodic 
level statistics, i.e., spectral or ensemble unfolding do not yield the same 
results~\cite{Asaga2001}. In addition to this, for $\beta=1$ the ensemble displays 
a large and robust quasi-degenerate portion of the spectrum for a wide interval of 
$k$, while for $\beta=2$ only seldom accidental quasi-degeneracies are 
observed~\cite{SHQ-Benet}. These results are due to the fact that each member of 
the ensemble is Liouville integrable in the semiclassical limit, for all values of 
$k$; see Ref.~\cite{BJL2003} for details. 

\section{Numerical results on fidelity decay}

\subsection{Definitions}
In order to study fidelity, as stated above, we must first define the reference (or 
unperturbed) and the perturbed 
Hamiltonians considered. The reference Hamiltonian $\hat{\cal H}_{0_k}$ shall 
be defined as the sum of a fixed one-body interaction term, and the diagonal part of the 
$k$-body interaction, with either $k=2$ or $k=3$. Since each of these terms have 
different spectral widths~\cite{Asaga2001}, we normalize each one with the 
respective width of the spectrum $W_k$. This mean-field unperturbed Hamiltonian 
is further restricted by imposing that $\hat{\cal H}_{k,0}$ is diagonal in the occupation 
number basis, which we shall use as our reference basis. Notice that the reference 
Hamiltonian contains the diagonal part of the $k$-body interaction, i.e., it includes the 
$k$-body matrix elements of the form $v_{r,r}^{(\beta)}$. Then, the unperturbed 
Hamiltonian is explicitly given by
\begin{equation}
\label{eq4}
\hat{\cal H}_{0_k}={1 \over{W_{k=1}}} \hat{H}_{k=1}^{\rm diag} + {\lambda \over{W_{k}}} 
\hat{H}_{k}^{\rm diag}.
\end{equation}
In \Eref{eq4}, the width of the spectrum of the $k$-body embedded ensemble is given 
by
\begin{equation}
  \label{width}
  W_k = \frac{1}{N}{\rm tr}\overline{[H_k^{(\beta)})]^2}
           = \Lambda_B^{(0)}(k)+{\delta_{\beta 1}\over{N}} \sum_{s=0}^k \Lambda_B^{(s)}(n-k),
\end{equation}
where
\begin{equation}
\Lambda_B^{(s)}(k) = {n-s\choose k}{n+s+1\choose k}.
\label{eqLambda}
\end{equation}
In these expressions we have used explicitly the restriction to the two single-particle 
level case ($l=2$); the label $B$ stands for bosons, the over-line indicates ensemble 
average, and $\Lambda_B^{(s)}(k)$ is the $s$-th eigenvalue of the 
ensemble-averaged correlation matrix of the bosonic $k$-body 
embedded ensemble; see Ref.~\cite{Asaga2001} for details.

\begin{figure}
  \includegraphics[height=0.32\textheight]{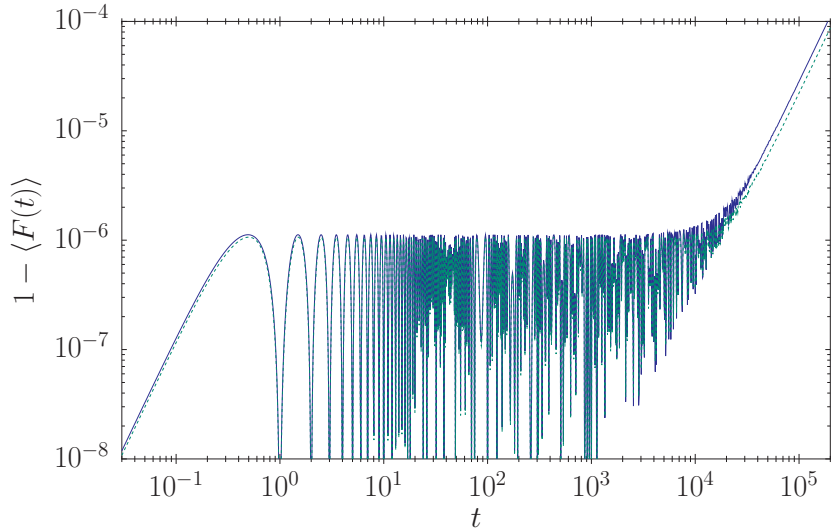}
  \caption{ Comparisson between the time evolution of the ensemble-averaged 
  fidelity illustrated with $1-\langle F(t)\rangle$ when the diagonal part of the 
  $k$-body interaction, which is included in the reference Hamiltonian 
  $\hat{\cal H}_{0_k}$, is fixed (upper curve for long times) or varied with the realizations 
  (lower curve for long times) for different realizations of the ensemble.}
  \label{fig0}
\end{figure}

As for the residual interaction, it simply consists of the remaining off-diagonal matrix 
elements of the $k$-body interaction, properly normalized by $W_k$. Denoting by 
$\lambda$ the perturbation strength, the perturbed Hamiltonian 
$\hat{\cal H}_{\lambda_k}$ is written as
\begin{equation}
\label{eq5}
  \hat{\cal H}_{\lambda_k} = \hat{\cal H}_{0_k} + {\lambda \over{W_{k}}} \hat{H}_{k}^{\rm off diag}.
\end{equation}

Notice that with this definition of the reference Hamiltonian and the traceless residual 
interaction, where $\hat{\cal H}_{0_k}$ also depends upon $\lambda$, the conditions 
to observe fidelity freeze are fulfilled~\cite{GKPSSZ2006}.
 
\subsection{Ensemble-averaged fidelity decay}
Here, we present numerical results for the ensemble-averaged fidelity decay in terms 
of the perturbation strength $\lambda$ and the number of particles $n$. For the 
numerical treatment, we have fixed the diagonal part of the $k$-body perturbation
to a specific realization of the ensemble. Therefore, the different realizations of the ensemble 
will only involve different realizations of the off-diagonal part of the perturbation. This
was done for numerical convenience. We emphasize that this implementation has 
no effect in the results; this is shown in Fig~\ref{fig0}, where the two curves included are 
difficult to distinguish quantitatively. In order
to obtain the physically relevant scalings, it is convenient to rescale the physical 
time $t'$ by a dimensionless time measured in units of the Heisenberg time, i.e., 
$t=t'/t_H$. Here, $t_H=2 \pi \hbar /d$ and $d$ denotes the average level-spacing 
of the spectrum of the unperturbed Hamiltonian $\hat{\cal H}_{0_k}$. 

The results for ensemble average were obtained as follows: We first fixed the 
unperturbed Hamiltonian $\hat{\cal H}_{0_k}$, and considered $1000$ 
independent realizations of the perturbation (involving only the off-diagonal
part of the $k$-body interaction); for a given perturbation strength $\lambda$ and fixing
the number of bosons $n$, this defined the perturbed Hamiltonian 
$\hat{\cal H}_{\lambda_k}$ through \Eref{eq5}. For each realization of the
residual interaction $\hat{H}_k^{\rm offdiag}$, we calculated the corresponding
time evolutions from an initial random state, which permitted to obtain the fidelity 
for this particular realization, using \Eref{eq2}. Averaging the resulting fidelities 
over different realizations constitutes the ensemble-averaged fidelity.

In \Fref{fig1} we present numerical results for the $1-\langle F(t)\rangle$ as a function of 
time (in Heisenberg time units), as convenient representation of ensemble-averaged 
fidelity decay, for various values of the perturbation strength $\lambda$ (for $n=1000$),
and various values of 
the number of particles $n$ (for $\lambda=10^{-6}$). The cases illustrated corresponds 
to $k=2$ and $\beta=1$; similar results were obtained for $k=3$ as well as 
when the time-reversal symmetry does not hold ($\beta=2$). As shown in \Fref{fig1}, 
for times smaller than the Heisenberg time $t<t_H=1$, but longer than Ehrenfest time, 
fidelity displays a quadratic decay in time, which is the typical situation observed for 
integrable systems. At  the Heisenberg time ($t=t_H=1$), the system exhibits a revival, i.e.,
$\langle F(t)\rangle$ approaches the unity again. This revival is natural for integrable
1d systems~\cite{Gorin2006}, which corresponds to this case, but is also 
observed for the Gaussian ensembles of Random Matrices~\cite{Stoeck2004}, or 
in models with underlying chaotic dynamics~\cite{Pineda2006}. 

\begin{figure}
  \centering
  \includegraphics[height=0.32\textheight]{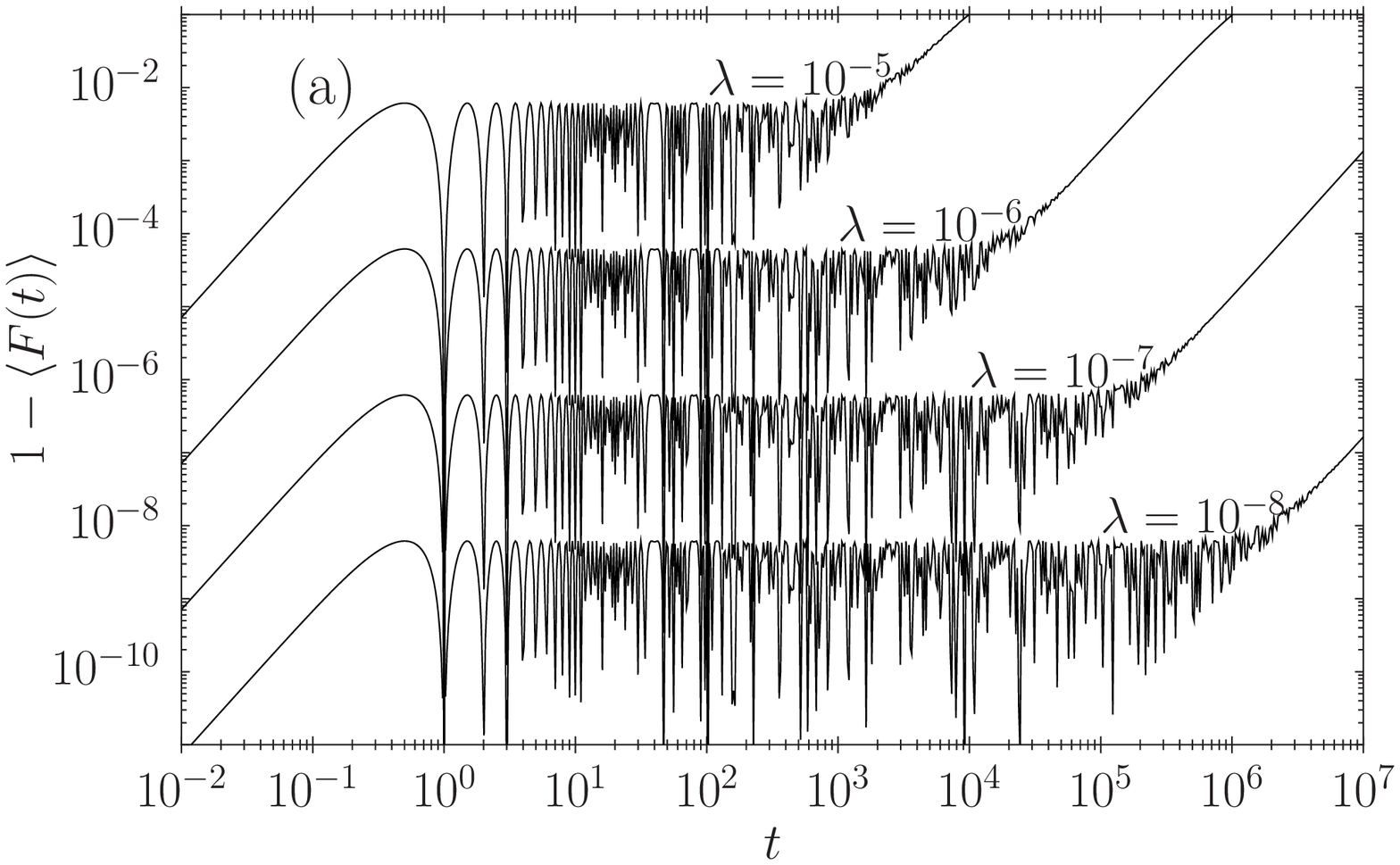}
\end{figure}
\begin{figure}
  \centering
  \includegraphics[height=0.32\textheight]{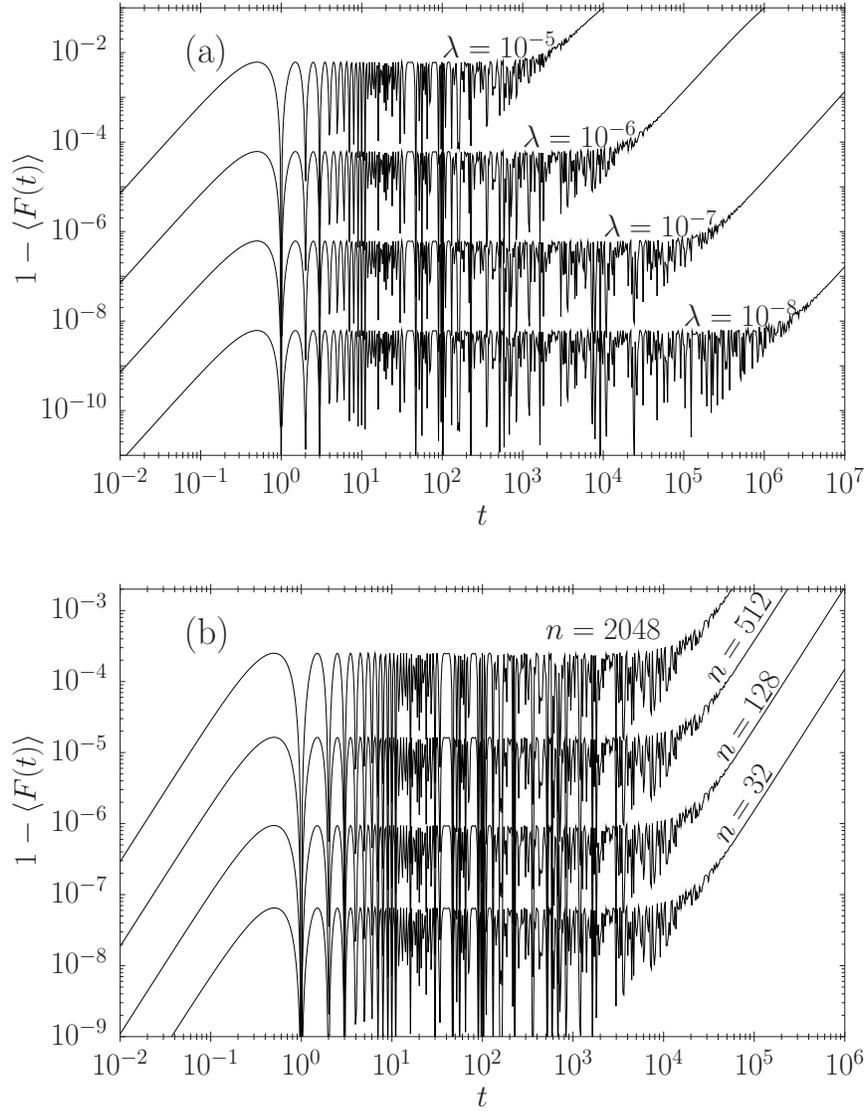}
  \caption{
    Time evolution of $1-\langle F(t)\rangle$ for $k=2$, $\beta=1$. 
    (a)~Dependence upon the strength of the perturbation ($n=1000$ particles); 
    (b)~dependence upon the number of particles for $\lambda=1\times 10^{-6}$. For very 
    short and very long times, the fidelity $\langle F(t)\rangle$ scales as $t^2$. 
    (c)~Illustration of the periodic revivals displayed by the ensemble-averaged 
    fidelity during the freeze. Note that the periodicity is the Heisenberg time.}
  \label{fig1}
\end{figure}

Beyond the Heisenberg time, fidelity remains essentially constant at certain 
value $F_{plateau}$; this is the fidelity freeze. The value $F_{plateau}$ scales 
as $\lambda^{2}$ and $n^{2}$ with respect to the perturbation strength and the 
number of particles, respectively. We note that during the freeze, fidelity displays 
some very short periodic revivals, whose period is precisely the Heisenberg time 
$t_H=1$. To the best of our knowledge, such periodic revivals during the freeze of
fidelity have not been observed. The freeze of fidelity lasts until some 
ending time $t_e$ is reached, which displays a $\lambda^{-1}$ dependence. 
After $t_e$, fidelity decays once again quadratically in time, until it saturates near $F=0$.

\setcounter{figure}{1}
\begin{figure}
  \centering
  \includegraphics[height=0.32\textheight]{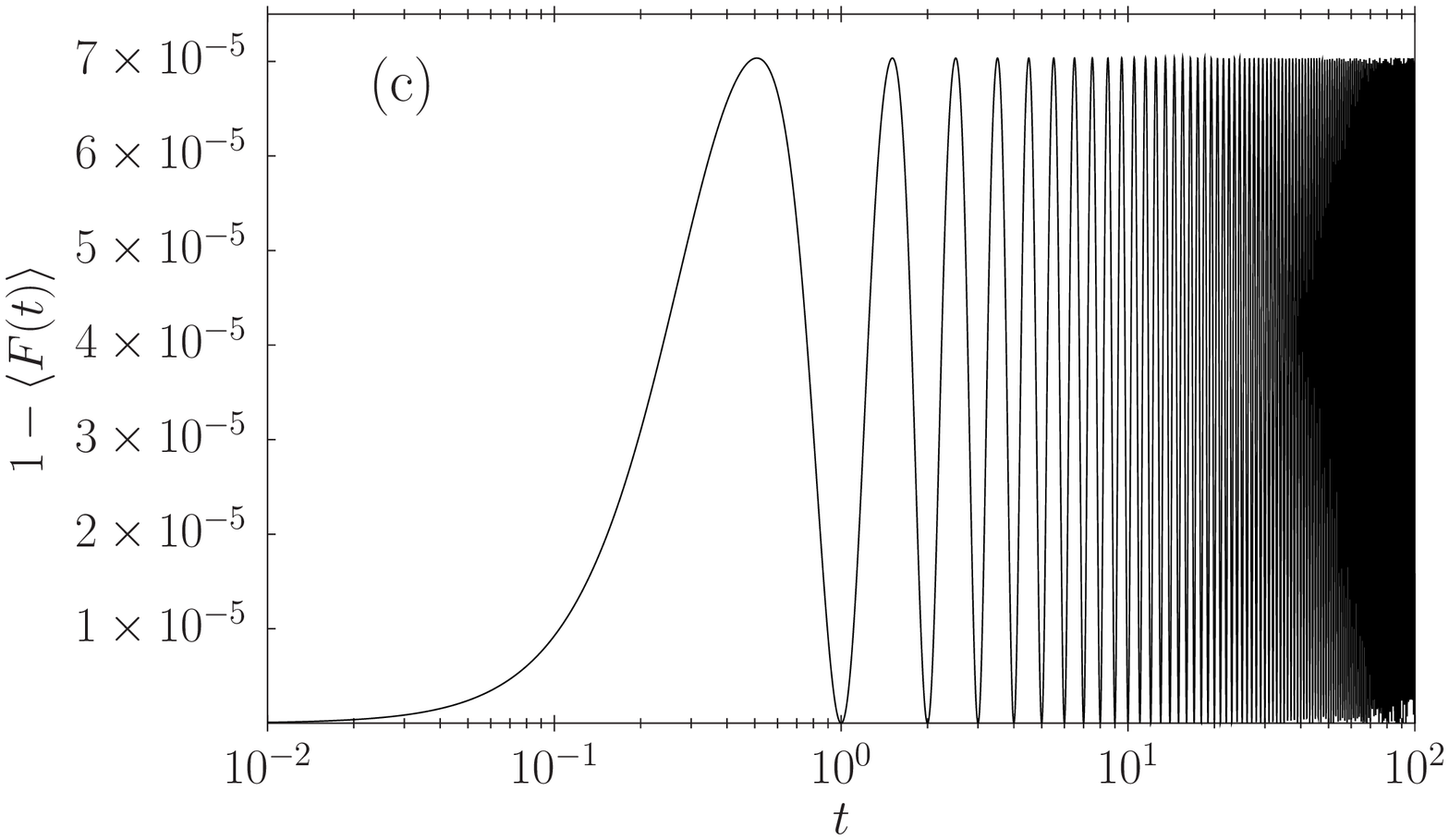}
  \caption{ (Continued) }
\end{figure}

\subsection{Fidelity decay of specific members of the ensemble}

Fidelity, for most of the realizations of the ensemble, resembles qualitatively 
the ensemble-average
fidelity. Yet, for some specific realizations the results do differ. A common case is that 
the position of the plateau may be slightly shifted up or down with respect to the 
average one. 

\begin{figure}
  \includegraphics[height=0.32\textheight]{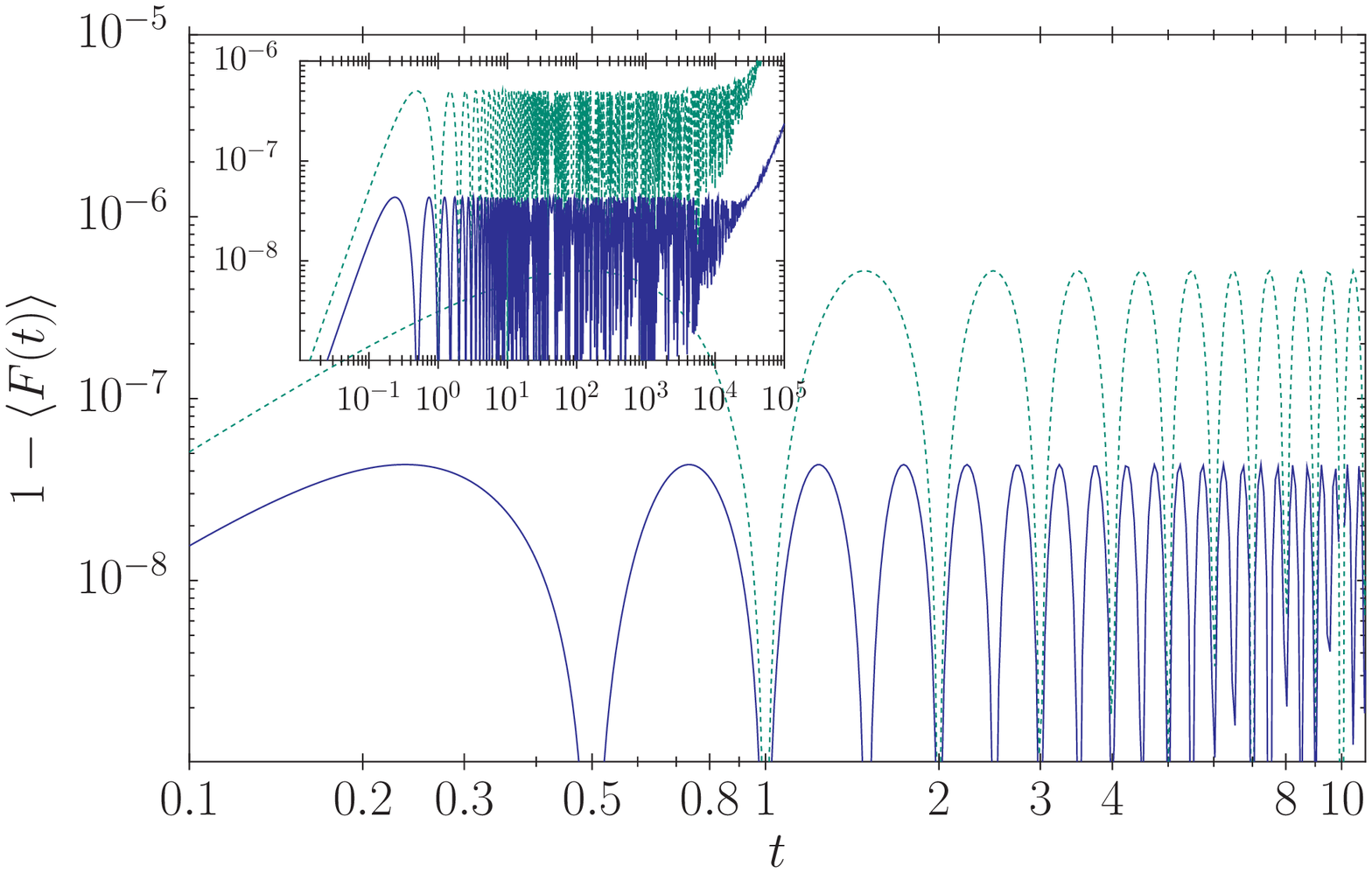}
  \caption{
    Fidelity decay for small times of two specific realizations of the ensemble
    for $k=2$ ($n=128$ and $\lambda=10^{-6}$). During the freeze of fidelity, 
    periodic revivals are displayed, with period $1$ (dotted curve), and of period $1/2$ 
    (solid curve), in units of the Heisenberg time. The inset presents the same results 
    for longer times.}
  \label{fig2}
\end{figure}

\begin{figure}
  \includegraphics[height=0.32\textheight]{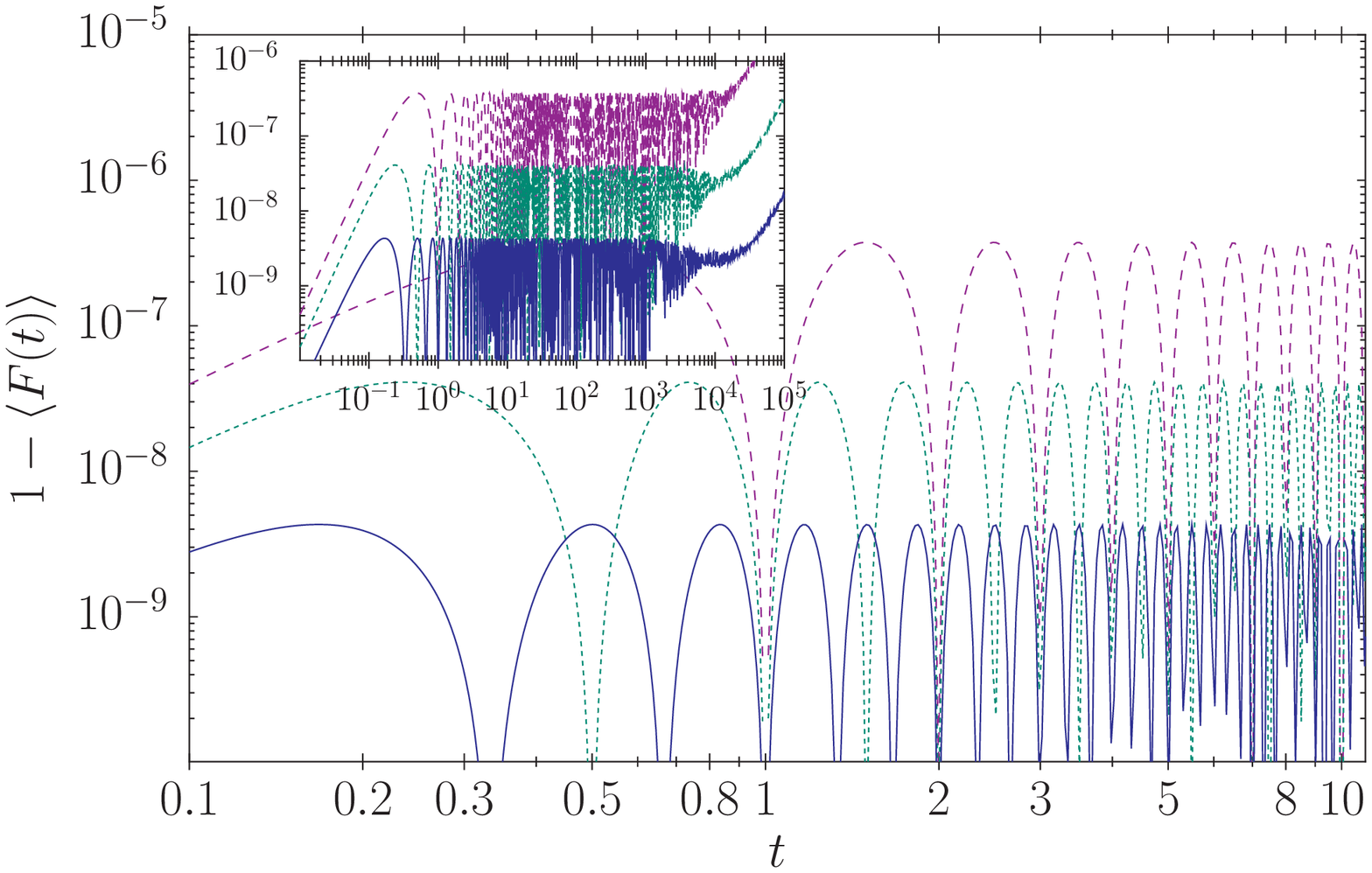}
  \caption{
    Same as \Fref{fig2} for $k=3$ ($n=128$ and $\lambda=10^{-6}$), displaying cases
    where the periodicity of the revivals is $1$ (dashed curve), $1/2$ (dotted curve) or 
    $1/3$ (solid curve), in units of the Heisenberg time. The inset shows the same results 
    for longer times.  }
  \label{fig3}
\end{figure}

A more interesting case occurs for some specific realizations of the ensemble, 
where the differences are more subtle and far reaching; see Figs.~\ref{fig2} and~\ref{fig3}. 
Indeed, for some specific members of the ensemble, fidelity displays 
periodic revivals during the freeze with an integer fractional period. That is, it displays a 
fractional period with respect to the unit, which is the 
Heisenberg time. This period has the form $T=1/c$, with $c$ an integer whose
specific value is linked with the value of the rank of the interaction $k$ of the 
residual interaction. 

These fractional periodic revivals are observed when some specific matrix elements of 
the $k$-body interaction matrix $v_{r,s}$ dominate over the rest. In fact, these matrix 
elements correspond to eliminating precisely $c$ particles from one single-particle
state, and creating them in the other one. By definition of the $k$-body interaction,
$c$ is bounded as $1\le c \le k$. Therefore, for $k=2$ we may move at most two particles
from one single-particle state to the other one. When the associated matrix elements 
somehow dominate the perturbation, fractional periodic revivals are observed with 
period $1/2$ in units of $t_H$ (see \Fref{fig2}). Likewise, for $k=3$ there are perturbing 
terms that involve moving up to one, two or three particles. The relative dominance of these 
terms is related to the observation of periodic revivals with periods $1$, $1/2$ or $1/3$ in 
Heisenberg-time units, respectively (see \Fref{fig3}). We note that these results do not
depend on the time-reversal invariance of the ensemble, given by $\beta$.

The results discussed above are illustrated in Figs.~\ref{fig2} and~\ref{fig3}. 
In~\Fref{fig2} we present the 
fidelity decay for short times (in units of $t_H$) of two realizations of the 
ensemble for $k=2$ ($n=128$ and $\lambda=10^{-6}$). The dotted curve displays
a case comparable with the ensemble-average fidelity, where the terms moving one 
particle dominate the perturbation; the solid curve illustrates
a case where the oscillations of period $1/2$ appear. In the latter case, the
dominating two-body interaction $\hat{H}_{k}^{\rm off diag}$ corresponds to the term 
$(\hat{a}_1^\dagger)^2 (\hat{a}_2)^2 + (\hat{a}_2^\dagger)^2 (\hat{a}_1)^2$. Figure~\ref{fig3} illustrates 
the $k=3$ case. As discussed above, in the case of three-body interactions
it is possible to observe oscillations of period $1$ (dashed line), $1/2$ (dotted line) 
or $1/3$ (solid line), depending on the specific $k$-body matrix elements that 
dominate the interaction.

\section{Conclusions}

In this paper we have presented numerical results on the fidelity decay of the two-level
bosonic $k$-body embedded ensembles of random matrices, for $k=2$ and $k=3$, 
considering the ensemble-averaged fidelity and comparing it with some individual 
realizations of the ensemble. The reference Hamiltonian is diagonal in the occupation 
number basis, where the residual interaction is traceless. We have obtained the relevant 
scaling laws, and observed the fidelity freeze, which displays oscillations of period 1 in 
units of the Heisenberg time.

We have also presented preliminary numerical results on the existence of  
fractional periodic revivals (in units of the Heisenberg time) during the freeze of fidelity,
for some specific members of this ensemble. Expressed in units of the Heisenberg time, 
the period of these revivals is $T=1/c$, where $c$ is an integer number
related to the specific $k$-body interactions which dominate the perturbation. The 
occurrence of these fractional periodic revivals is somewhat rare with respect to 
ensemble average;  we conjecture that this is related to the relative number of terms 
in the perturbation which move $c$ particles from one level to the other. Our results
indicate that the appearance of the fractional periodic revivals is independent of the
time-reversal invariance. These results may be interesting for the understanding 
---and even the measurement--- of three-body interactions in two-component 
Bose-Einstein condensates~\cite{GatiOberthaler2007}.

\begin{theacknowledgments}
We acknowledge financial support from the projects IN-114310 (DGAPA-UNAM) 
and 57334-F (CONACyT). The Hungarian-Mexican Intergovernmental 
S \& T Cooperation Program (MX-16/2007, NKTH, and I0110/127/08, CONACyT) is
also acknowledged. LB thanks the kind hospitality of \`A. Jorba and C. Sim\'o at 
the U. Barcelona, where this work was completed.
\end{theacknowledgments}

\bibliographystyle{aipproc}   

\end{document}